\documentclass{style/nseJournal}

\begin{document}

\title{Production of high purity $^{52g}$Mn from $^{nat}$V targets with $\alpha$ beams at cyclotrons}

\addAuthor{Colombi A. }{a,b}
\addAuthor{Carante M.P. }{a}
\addAuthor{Barbaro F. } {c}
\addAuthor{Canton L. }{c}
\addAuthor{\correspondingAuthor{Fontana A. }}{a}
\correspondingEmail{andrea.fontana@pv.infn.it}
\addAffiliation{a}{INFN - Sezione di Pavia} 
\addAffiliation{b}{Dipartimento di Fisica, Universit\`a di Pavia} 
\addAffiliation{c}{INFN - Sezione di Padova} 

\addKeyword{Cyclotron radionuclide production}
\addKeyword{$^{52g}$Mn}
\addKeyword{$^{53}$Mn}
\addKeyword{$^{54}$Mn} 
\addKeyword{multi-modal imaging} 
\addKeyword{$\alpha$-induced reactions} 
\addKeyword{nuclear reactions modeling}

\titlePage 

\begin{abstract} % abstract
Radioisotope $^{52g}$Mn is of special interest for multimodal imaging. Using state-of-art nuclear reaction codes, 
we study the alternative nuclear reaction route  $^{nat}$V($\alpha$,x)$^{52g}$Mn in comparison with the standard production routes based upon the use of chromium targets.   
The integral yields of $^{52g}$Mn and contaminants have been evaluated. The main outcome of this investigation is that the production of the main contaminant isotope $^{54}$Mn is expected to be lower than with $^{nat}$Cr.
The study also reveals a large spread in the cross-section data set and points out the need of more precise measurements of the reaction  $^{nat}$V($\alpha$,x)$^{52g}$Mn as well as the need of a more accurate theoretical description.
\end{abstract}

\section{Background}

The search and production of innovative radioisotopes for medical applications is a topic of great interest nowadays,  particularly  for advancements in theranostics and multimodal imaging.
In this work, we are interested in the latter case, which boosts up the diagnostic image information by simultaneously using two different physical processes. 
For example, to obtain a combined PET/MRI scan using a single, manganese-labelled, molecular agent it is necessary to first produce a $^{52g}$Mn-labelled compound for PET and, separately, the same compound containing only paramagnetic manganese for MRI. These two molecular agents should have the same chemical composition in order to probe the same biological vector, but they act separately because of the differences in sensitivity requested by PET and MRI scans. Manganese radionuclides suitable for PET scan are $^{52g}$Mn, $^{52m}$Mn, and $^{51}$Mn \cite{brandt2019}.
We are interested here in the production of $^{52g}$Mn: 
its decay properties, and those of contaminants typically involved in the production methods, are reported in Tab. \ref{tab:1}.

The radionuclide $^{52g}$Mn appears of particular interest because of its relatively long half-life (5.6 d) suitable for the radiolabeling of antibodies and other slow biological compounds for the study of pharmacokinetics.
Its $\beta^+$ emission is characterized by a very low (i.e. endpoint) energy, about 0.6 MeV, thus leading to a very good resolution of PET scans. On the other hand, $^{52g}$Mn decay to $^{52}$Cr occurs with the emission 
of three prompt $\gamma$ rays (with energies $744.2, 935.5$ and $1434.1$ keV)
which provide an additional contribution to the patient's dose increase in clinical applications and could interfere with erroneous signals in PET's $\gamma$-based image  reconstruction.

The effective dose burden due to the use of $^{52g}$Mn 
as brain tracer under the simple MnCl$_2$ chemical compound has been carefully investigated with computational dosimetry codes in Ref. \cite{denardo2019}. It was found that
the radiation dose released by $^{52g}$Mn has been estimated to be about 130 times the dose released by $^{51}$Mn. Since the manganese-chloride compounds are retained in the body for a long time ($i.e.$ 4.5 days for the fast component and 44.6 days for the slow component)\cite{mahoney1968}, the longer physical half-life of the $^{52g}$Mn radionuclide affects negatively the effective dose imparted to the patient. 

The standard cyclotron-based production of $^{52g}$Mn 
relies on the following nuclear reaction routes: $^{52}$Cr(p,n)$^{52g/m}$Mn,
$^{53}$Cr(p,2n)$^{52g/m}$Mn, and \\
$^{54}$Cr(p,3n)$^{52g/m}$Mn, 
which can be exploited at proton energies within 20 MeV or, even, with 16 MeV cyclotrons \cite{brandt2019}. The use of natural Chromium targets has been extensively investigated in the past: $^{52}$Cr constitutes 84\% of $^{nat}$Cr, while $^{53}$Cr is 10\% and  $^{54}$Cr is 2\%; the remaining 4\%, made of $^{50}$Cr, contributes only to contaminants production but not to the production of the radionuclide concerned. 
 The cross section for the proton driven reaction route
 on $^{nat}$Cr is large enough to provide sufficient yield for pre-clinical applications.
Production with natural Chromium targets is favourable also because the radionuclidic purity is very high  \cite{Lapy2019} for long periods (up to a few months), thanks also to the long half-life of $^{52g}$Mn.

The main drawback due to the use of a natural target is the expected yield of impurities during the irradiation, 
in particular the long-lived $^{53}$Mn and $^{54}$Mn radionuclides, mainly via the reaction routes:
$^{53}$Cr(p,n)$^{53}$Mn,
$^{54}$Cr(p,2n)$^{53}$Mn, and
$^{54}$Cr(p,n)$^{54}$Mn.
The production of both long‐lived
contaminants can be avoided by irradiation
of highly enriched $^{52}$Cr targets and should therefore not impose
stringent limitations for potential clinical uses of $^{52g}$Mn. However the use of enriched material  significantly increases the production costs and implies the development of an efficient and cost-effective target material recovery protocol.

\begin{table}[!htb]
\renewcommand*{\arraystretch}{1.5}
\caption{Decay characteristics of the manganese radioisotopes (IT: Isomeric Transition, EC: Electronic Capture). Radioactive decay products are indicated with an asterisk.}
\label{tab:1}
\centering
\begin{tabular}{|ccccc|}
\hline
Radionuclide & Half-life & Decay mode & Branching ratio  & Daughter  \\
\hline
$^{51}$Mn & 46.2 m & $\beta^+$ & 100\% & $^{51}$Cr$^*$  \\
$^{52g}$Mn & 5.6 d & $\beta^+$ & 100\% &  $^{52}$Cr   \\
$^{52m}$Mn & 21.1 m & $\beta^+$ & 98.25\% &  $^{50}$Cr   \\
$^{52m}$Mn & 21.1 m &  IT & 1.75\% &   $^{52g}$Mn$^*$ \\
$^{53}$Mn & $3.6\times10^6$ y & EC & 100\% &  $^{53}$Cr   \\
$^{54}$Mn & 312 d & EC & 100\% &   $^{54}$Cr  \\ 
\hline
\end{tabular}

\end{table}

The main purpose of this study is therefore the search for an alternative and competitive route to produce 
$^{52g}$Mn with high radionuclidic purity and high production yield. 
This work has been carried out in support to the activities of the METRICS (Multimodal pET/mRi
Imaging with Cyclotron produced $^{51/52}$Mn and stable paramagnetic Mn iSotopes) project in the framework of the SPES/LARAMED research program at INFN-LNL.

As a promising radionuclide production pathway, the nuclear reaction route $^{nat}$V($\alpha$,x)$^{52g/m}$Mn, which is dominated by $^{51}$V($\alpha$,$3n$)$^{52g/m}$Mn, has been investigated. 
Generally, this possible production route is neither mentioned in the medical radioisotope literature (see for a recent review \cite{Qaim2020}), nor 
in the IAEA Livechart website \cite{livechart}, an internationally recognized reference for the production of medical radioisotopes.
The IAEA Livechart website mentions helium beams only for $^3$He particles in the generator-like sequence: 
$^{52}$Cr($^3$He,3n)$^{52}$Fe to $^{52}$Fe(EC \, $\beta^+$)$^{52}$Mn. 

The $^{nat}$V($\alpha$,x)$^{52}$Mn reaction route has been studied by various authors in the last 50 years, both for the ground and for the metastable state. The experimental data for this reaction are collected in the EXFOR database \cite{EXFOR}. 
Different routes to yield the $^{52}$Mn isotope have been measured and reported in \cite{Dmitriev1969} based upon the use of natural targets with Cr, V and Fe bombarded with proton, deuteron or alpha beams.
Both cross-sections and integral yields are given with an estimated error of 13\%. A first attempt to compare theory models and experimental results with $\alpha$-particle beams on a variety of targets, including $^{51}$V, is given in  Ref.\cite{Bowman1969}.
Several measurements have thus been performed with important advances in the experimental techniques and compared with reaction models (statistical and pre-equilibrium mechanisms) that have been gradually refined and improved over the years: \cite{Michel1982},  \cite{Rama1987}, \cite{West1987}, \cite{Sonzogni1993}, \cite{Ismail1993}, \cite{Singh1995}, \cite{Chowdhury1995}, \cite{Bindu1998}, \cite{Peng1999}.
Finally, new results on alpha particles on Vanadium targets have been published recently,
in \cite{Ali2018}.

Overall, the measured excitation functions show a similar structure with an initial peak followed by a decrease: the peak corresponds to alpha emission mainly due to the evaporation process of the compound nucleus, while the tail is dominated by the pre-equilibrium decay. The spread of collected data is significant and can be attributed to the long period over which the data were taken and to the different experimental techniques used over the years. The large number of published measurements and the large spread in the data demands for an accurate nuclear data evaluation of the cross sections, given its importance for medical applications.
In addition, it is important to assess also the theoretical uncertainty arising from the different nuclear models employed and this will be taken into account in the present work.

To identify the energy intervals and irradiation conditions most suited for the radionuclide production, different nuclear model tools have been used.
After performing the excitation functions analysis, the thick target  yield for a hypothetical irradiation with $\alpha$ particles, on a $^{nat}$V target of a given thickness, has been calculated and from there the time-evolution of  the related isotopic and radionuclidic purities, assuming a sufficiently long cooling time.

At last the results obtained by using $^{nat}$V targets and $\alpha$ beams have been compared with those derived 
from natural Chromium with proton beams. In addition, a comparison with enriched Chromium targets, considering both proton and deuteron beams, is provided.

\section{Methods} \label{codes}

\subsection{Nuclear model calculations}
\label{codes2}
The study of the aforementioned highlighted nuclear reaction routes implies the adoption of different 
models to describe both the compound nucleus formation/decay and pre-equilibrium dynamics.
To this purpose three of the most up-to-date nuclear reaction codes: Talys \cite{talys}, Empire \cite{empire} and Fluka \cite{batt2007} have been used.
The nuclear reaction mechanisms relevant for radionuclide's production at cyclotrons are dominated by the
compound nucleus formation and by pre-equilibrium emission and all the three codes are based upon the  
nuclear reaction models developed to describe these processes. A quick review about all nuclear models used 
can be found in \cite{pupillo2019} and in the codes's references.

Talys (version 1.9) is a software for the simulation of nuclear reactions that includes many 
state-of-the-art nuclear models to cover most of the reaction mechanisms encountered in light 
particle-induced nuclear reactions \cite{talys}. 
The nuclear reaction rates evaluated by the code are based upon the Hauser-Fesbach model \cite{hauser} for the equilibrium mechanisms and on four different theoretical frameworks for the pre-equilibrium process.   
The level density is another important aspect to consider for describing the reaction and Talys has six possible options for its description, ranging from the simplest Fermi gas model to more complex
microscopic approaches. 

Empire (version 3.2.3) is a nuclear reaction code based on various
nuclear models and designed for calculations over a broad range of energies (from
a few keV up to hundreds MeV) and various incident particles (nucleons, photons, deuterons
and light ions). The code accounts for the major current nuclear reaction
models, such as Optical Models, Coupled Channels and DWBA (Distorted Wave
Born Approximation) models for elastic and inelastic scattering; Exciton model
and Hybrid Monte Carlo Simulations for pre-equilibrium emission; and finally the
Hauser-Feshbach model for compound nucleus \cite{empire}.

Fluka (development version 2018.2) is a general purpose code for modelling particle transport 
and interaction with matter; 
it covers an extended range of applications, spanning from proton and electron accelerator 
shielding to calorimetry, dosimetry, detector design, radiotherapy and more \cite{batt2007,cascad,inf2015}. 
The code, based on the PEANUT (PreEquilibrium Approach to Nuclear Thermalisation) module,
can be used to calculate the production of residual nuclei and, in many cases, results have already been validated with experimental data. Residual nuclei (and, thus, radionuclides) emerge
directly from the inelastic hadronic interaction models and can be calculated 
for arbitrary projectile-target configurations (including
nucleus-nucleus interactions) and energies. Regarding the production of isomers, 
the Fluka version used in this work does not have a built-in routine to predict the correct branching 
for the production of different states of the same radionuclide, 
but it distributes the cross section equally over the different states: for this reason, we only consider 
the results of Fluka in the cases where the separation between ground and metastable nuclides is not explicitly involved.

\subsection{Uncertainty evaluation with Talys}
A total of 24 different model combinations may be available with Talys by taking advantage of the different possible level density and pre-equilibrium options.
Many calculations found in literature report the use of 
a default option, however this is not always the best choice and therefore alternative option configurations have been introduced and evaluated. As an exemplum giving, the so-called ``Talys adjusted" configuration reported in Ref. \cite{Duchemin:2015} has been often used.
However, both cases rely on the selection of a single model for level density and pre-equilibrium, disregarding all the others, and not exploiting the full potentiality and  versatility of the code.
In the next, we introduce a novel way to deal with the theoretical 
variability provided by the different models. Instead of plotting all 
the 24 curves, we compare the different models introducing a 
statistical band, along with similar ideas explored in Ref. \cite{Lamere2019}. Starting from the 24 different model calculations, a band is constructed from
the interquartile range, given by the difference between  the third ($Q_3$) and the first ($Q_1$) quartile. In addition we introduce for each energy a ``Best Theoretical Evaluation" (BTE) of the cross section 
by taking the average of the first and third quartile, and associate 
to it the uncertainty given by the half-width of the interquartile band:

\begin{equation}\label{eq:1}
	\sigma_{BTE}=\frac{Q_1+Q_3}{2}, \qquad
	\Delta\sigma_{BTE}=\frac{Q_3-Q_1}{2}.
\end{equation}
Some models of the ensemble may show a too large variability, over- or  under-estimating by large the data. If we consider the interquartile band, 
spanning quartiles $Q_1$ and $Q_3$, only the central 50\% calculations are retained, and this leads to a more reasonable description.

In this way a reference value for cross-section depending upon all models provided by the code, and a statistical uncertainty depending upon the variability of the models themselves, may at last be obtained.
The same procedure has been applied to get an assessment not only for cross sections, but also yields, activities, isotopic, and
radionuclidic purities, both for the radionuclide concerned and as well as its main contaminants.
The BTE approach derives from descriptive statistics and connects to the concept of trimmed average: it provides a robust estimator of the theoretical cross section and allows to discard, 
in a consistent way, the outlier values provided by a subset of the 
models. These outliers are shown, for instance, in Fig. 1 with the 
two dashed lines, corresponding to the maximum and minimum values provided by the 24 model calculations. 
This approach is alternative to other, more sophisticated techniques, developed to
introduce a theoretical uncertainty band, like for example the multistep 
method \cite{Hussain:2010}, in which a rescaling of the models to experimental 
data is performed, or total Monte Carlo techniques \cite{TENDL}, in which 
the parameters of the models are sampled randomly to assess the variability 
of the calculation outcomes. Our description appears more practical, since it trims the calculations 
at the edge of the set and provides quickly the final result in a single 
deterministic step. 

\section{Results}
\subsection{Cross sections analysis} \label{Crosssections}

%QUI DA RISCRIVERE MEGLIO

\begin{figure}[!htb]
\centering
\includegraphics[width=12cm]{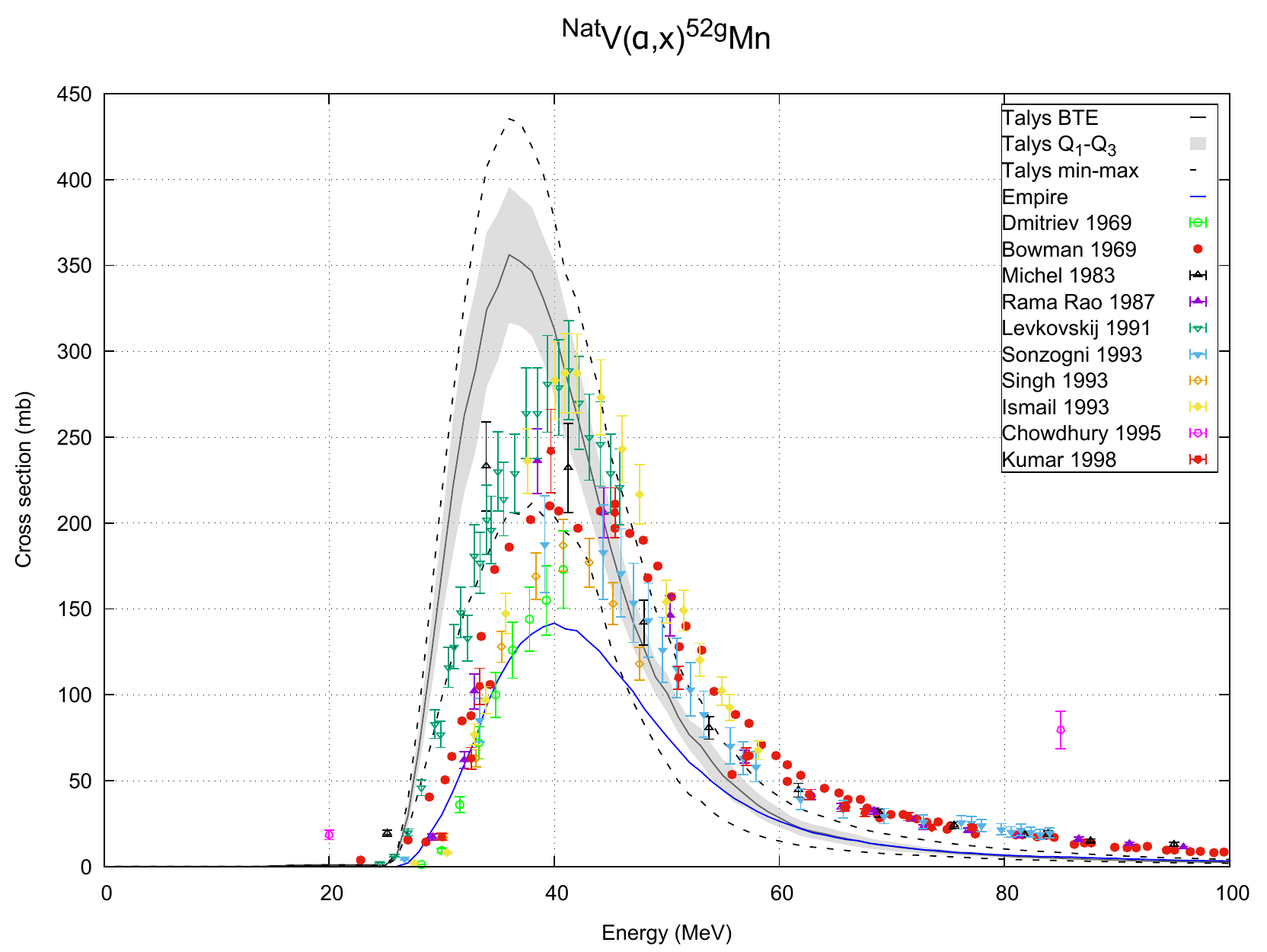}
\caption{Cross section for $^{nat}$V($\alpha$,x)$^{52g}$Mn route as predicted 
by three different codes and compared with the experimental data currently available in the EXFOR 
database \cite{EXFOR}. To take into account the theoretical uncertainty of the all the models available in Talys, a grey band for the quartiles $Q_1$-$Q_3$ and two dashed line for the minimum and maximum are plotted.}
\label{fig:1}
\end{figure}

The cross section of the nuclear reaction route  $^{nat}$V($\alpha$,x)$^{52g}$Mn is plotted in Fig. \ref{fig:1}. The experimental data taken from the available databases (EXFOR) are compared with the calculated results obtained with the reaction codes Talys, Empire, and, when relevant, Fluka (see SubSect. \ref{codes2}).  A dispersion of data, accumulated over a period of few decades, may be clearly seen, thus complicating the precise evaluation of the production yield. Talys results are shown following the scheme discussed in section \ref{codes} to take into account the variability of the models: a ``best theoretical evaluation" (solid line), 
an interquartile range (gray band), 
and the min and max values resulting from all the models considered (dashed lines). Unfortunately, even taking into account the variety of Talys models, in the energy region below 45 MeV, calculations overestimate significantly the trend of data, however the spread of data prevents a precise determination of the overestimation factor.  
Likewise, Empire calculations are slightly lower than the measured cross section, but the extension of the underestimation is difficult to be evaluated. 
As it will be shown in the following, the problems with the theoretical description seem to be restricted to this particular channel; indeed the description of the other production routes or the contaminants cross-sections turn out to be in better agreement. In this work we do not provide a solution to this problem, e.g. by an attempt to improve the theoretical models. Nevertheless, the analysis of yields and purities performed from various sources in the second part of this work (see Tab. \ref{tab:3}) are deemed sufficient to consider this route of interest.

\begin{figure}[htb!]
\centering
\includegraphics[width=12cm]{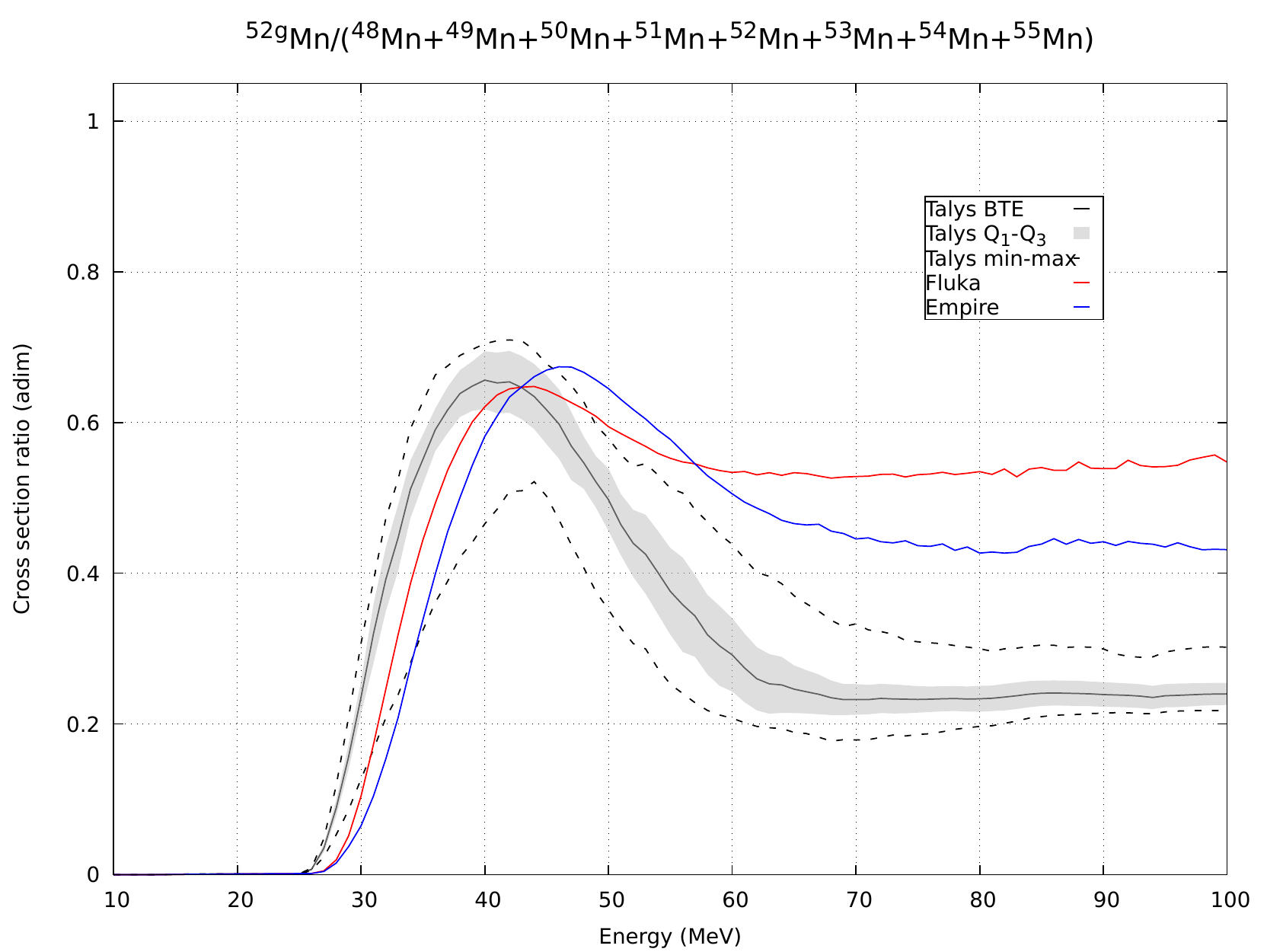}
\caption{Ratio of the calculated cross sections for $^{nat}$V($\alpha$,x)$^{52g}$Mn and total of other Mn isotopes.}
\label{fig:2}
\end{figure}

The cross-sections ratio between $^{52g}$Mn and the sum of all Mn contaminants expected cross sections is shown in Fig. \ref{fig:2}, in order to determine the optimal energy region where such a quantity is as high as possible. Based upon the codes expected cross-section trends, the resulting maximum for such a ratio turns out to be close to the maximum of the cross section at 40 MeV. Performing an irradiation in an energy range around such a value would thus lead to an high as possible radionuclide quality for the $^{52g}$Mn, having the minimum expected level of contamination. This residual contamination, however, would not be negligible, since the cross section ratio has a maximum value of about 0.6.

Nevertheless, it is important to observe that the majority of the produced isotopes are characterized either by a very short, or by a very long, half-life. In particular, $^{48}$Mn, $^{49}$Mn, $^{50g/m}$Mn, $^{51}$Mn and $^{52m}$Mn have half-lives smaller than one hour and their contamination, both in terms of isotopic and radionuclidic purity, are thus negligible just after a few hours. $^{55}$Mn is stable and does not affect the radionuclidic purity. Moreover, it is produced in the electromagnetic channel with very low excitation function. Also, $^{53}$Mn, whose half-life is of about 3.6$\times10^6$ years, does not affect the radionuclidic purity in a significant way and does not release a significant dose to the patient. Finally $^{54}$Mn,  with an intermediate half-life of about 312 days, could represent an issue in terms of the expected dose increase to the patient.

\begin{figure}[htb!]
\centering
\includegraphics[width=12cm]{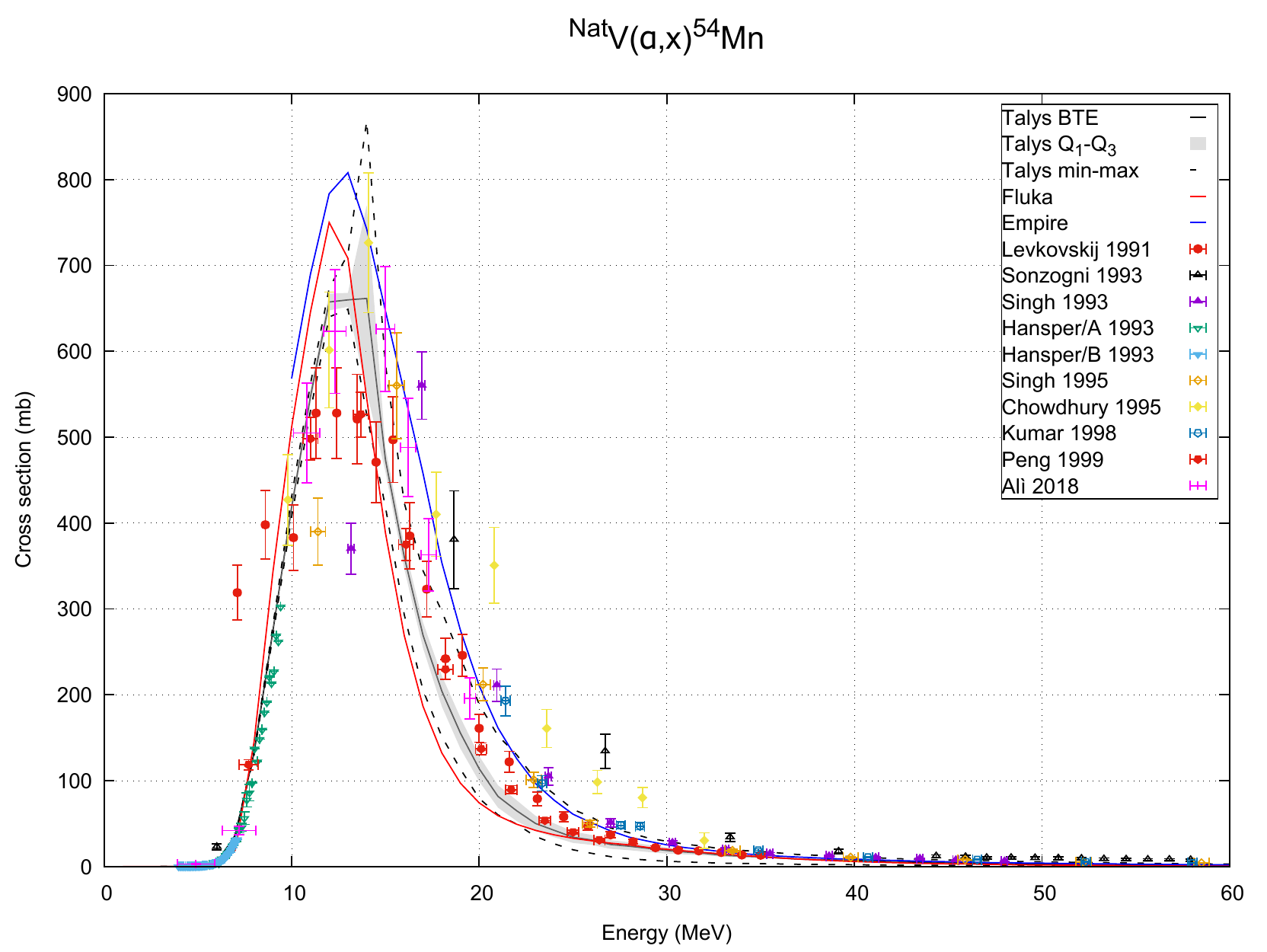}
\caption{Cross section for $^{nat}$V($\alpha$,x)$^{54}$Mn. The meaning of the theoretical lines and bands is the same as in Fig. \ref{fig:1}.}
\label{fig:3}
\end{figure}

For this reason, in Fig. \ref{fig:3} the cross section of the reaction $^{nat}$V($\alpha$,x)$^{54}$Mn is shown. The agreement between experimental data (\cite{Sonzogni1993}, \cite{Singh1995}, \cite{Chowdhury1995}, \cite{Bindu1998}, \cite{Peng1999}, \cite{Ali2018}, \cite{Levkovskij1991}, \cite{Singh1993}, \cite{Hansper1993}) and the three nuclear codes is satisfactory, especially if compared to Fig. \ref{fig:1}. Still, there are about 25\% differences among the various calculations around the peak (at 13 MeV) which is not far from a 30\% variability of the experimental data in the same energy region. On the contrary, at the maximum production of $^{52g}$Mn, around 40 MeV, the cross section for $^{54}$Mn is very low. This fact is plain if we take into account only $^{54}$Mn as contaminant and we plot the quantity

\begin{equation}
	r=\frac{\sigma_{^{52g}\mbox{\rm Mn}}}{\sigma_{^{52g}\mbox{\rm Mn}}+\sigma_{^{54}\mbox{\rm Mn}}},
\end{equation}

\begin{figure}[htb!]
\centering
\includegraphics[width=12cm]{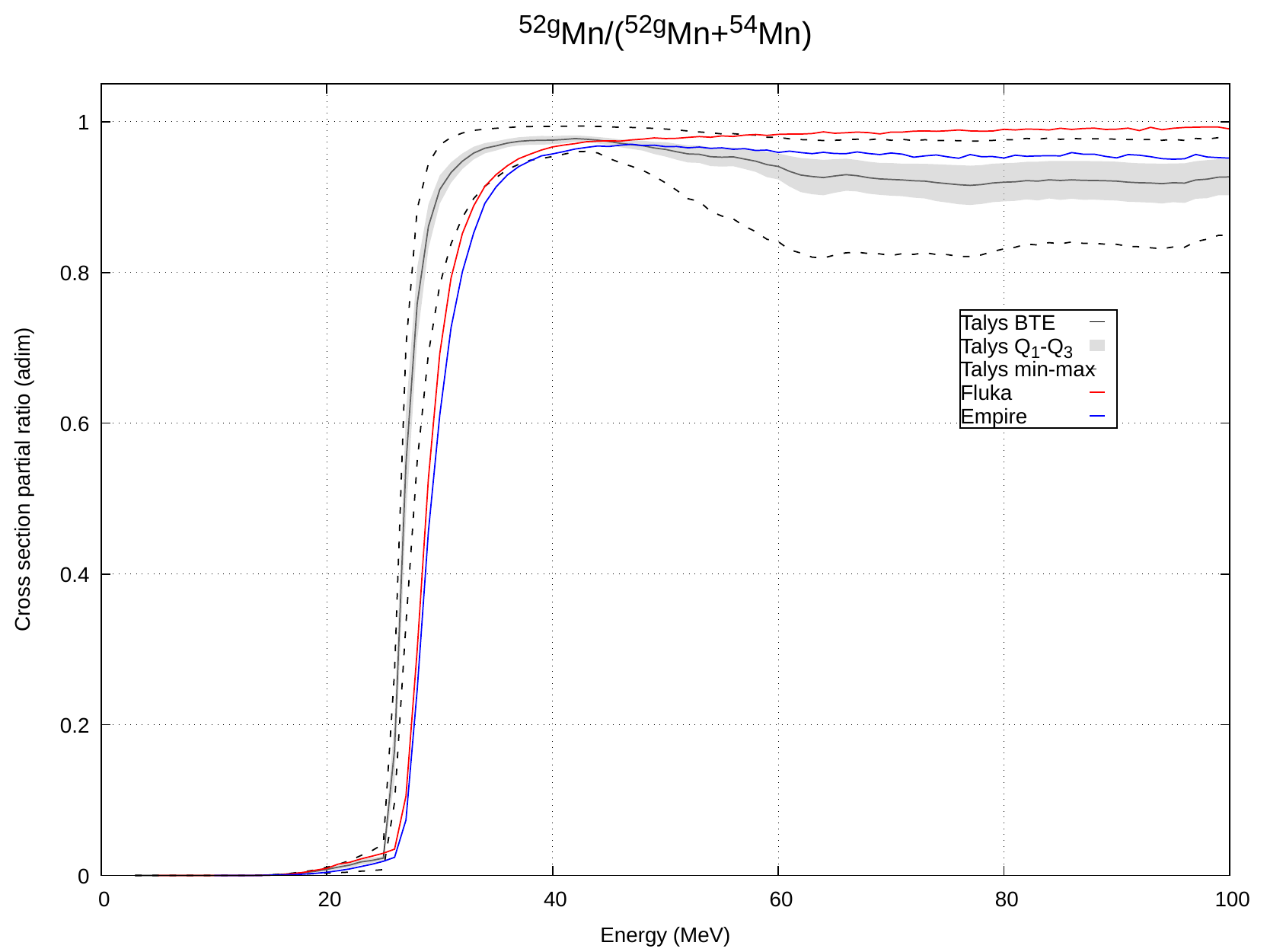}
\caption{Partial ratio of the calculated cross sections for $^{nat}$V($\alpha$,x)$^{52g}$Mn and the 
$^{54}$Mn isotope. The energy window with the possibility of high purity production of $^{52g}$Mn due to the 
favorable interplay of the different reactions thresholds is clearly visible.}
\label{fig:4}
\end{figure}

as shown in Fig. \ref{fig:4}. Above the energy of about 30 MeV, the production of $^{52g}$Mn is almost pure, with respect to its most dangerous contaminant. In the next Section we will focus on the region around 40 MeV, to evaluate the $^{52g}$Mn production yields and purities.

\subsection{Yields and purities}

Once the most convenient energy window for the production of $^{52g}$Mn is identified, it is possible to plan the optimal irradiation conditions. The production rate of a nuclide for a beam impinging on a target of a given material and thickness can be calculated with the formula \cite{IAEAreport,Iliadis}

\begin{equation}
	R = \frac{I_0}{z_{proj}|e|}\frac{N_a}{A}\int_{E_{out}}^{E_{in}}\sigma(E)\left(\frac{dE}{\rho_tdx}\right)^{-1}dE \, ,
\end{equation}

where $I_0$ is the charge per unit time hitting the target, $z_{proj}$ the charge state of the incident particle (2 in the case of a completely ionized $^4He$ beam), $e$ the electron charge, $N_a$ the Avogadro number, $A$ the target atomic mass, $E_{in}$ and $E_{out}$ the energy of the projectile impinging on the target and after exiting from the target, respectively, $\sigma(E)$ the production cross section for the nuclide, $\rho_t$ the target density and dE/dx the stopping power of the projectile in the target, calculated with the Bethe-Bloch formula \cite{leo}. In this case the irradiation parameters are: beam current of 1 $\mu A$, incident energy of 48 MeV, target thickness of 200 $\mu m$ (corresponding to $E_{out}=33.9$ MeV), and irradiation time of 1 h.  Radionuclides from $^{50}$Mn to $^{55}$Mn are produced in this energy window.

The yield rate for all the Mn isotopes of interest are calculated, and for $^{52g}$Mn it was found to be between 5.6$\times10^8$ and 1.45$\times10^9$ nuclei$\cdot$s$^{-1}$, depending upon the different codes and models. From the rate, the time evolution of the number of nuclei of a specific isotope can be obtained, during and after the irradiation, by means of standard Bateman equations. Every manganese radionuclide of interest decays in different chemical elements, with the only exception of $^{52m}$Mn, which decays in $^{52g}$Mn with a branching ratio of 1.75$\%$ and with a half-life of about 21.1 minutes.

\begin{figure}[htb!]
\centering
\includegraphics[width=12cm]{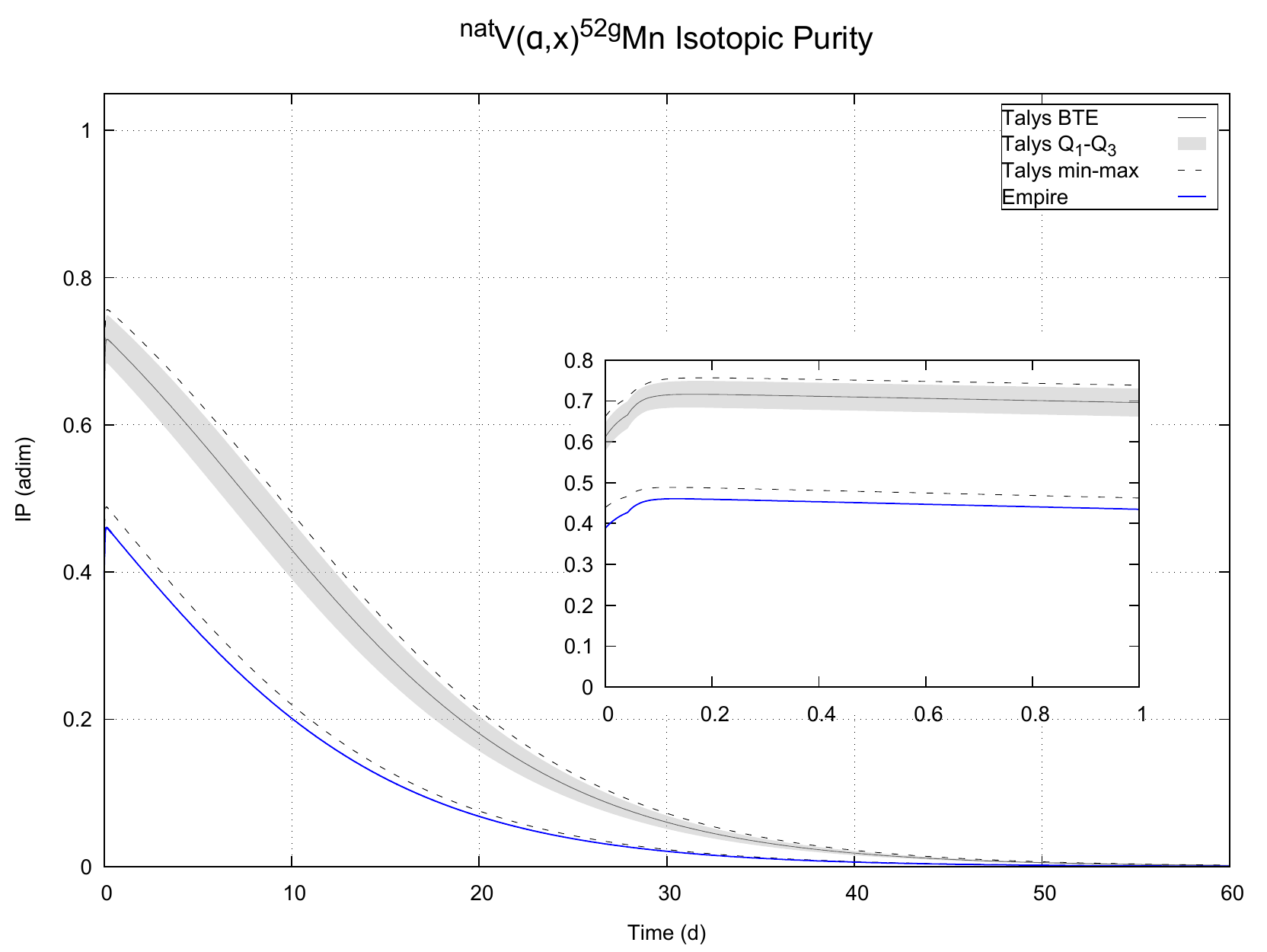}
\caption{Post-irradiation time evolution of the $^{52g}$Mn Isotopic Purity (IP) expected by nuclear models for a long time window with an hypothetical one hour irradiation of a $^{nat}$V target with a beam energy of 48 MeV, thickness 200 $\mu m$ (corresponding to an exit energy of 33.9 MeV), current of 1 $\mu A$.  The inset shows the evolution for the first 24h, including the irradiation (i.e. build-up) time.}
\label{fig:5}
\end{figure}

\begin{figure}[htb!]
\centering
\includegraphics[width=12cm]{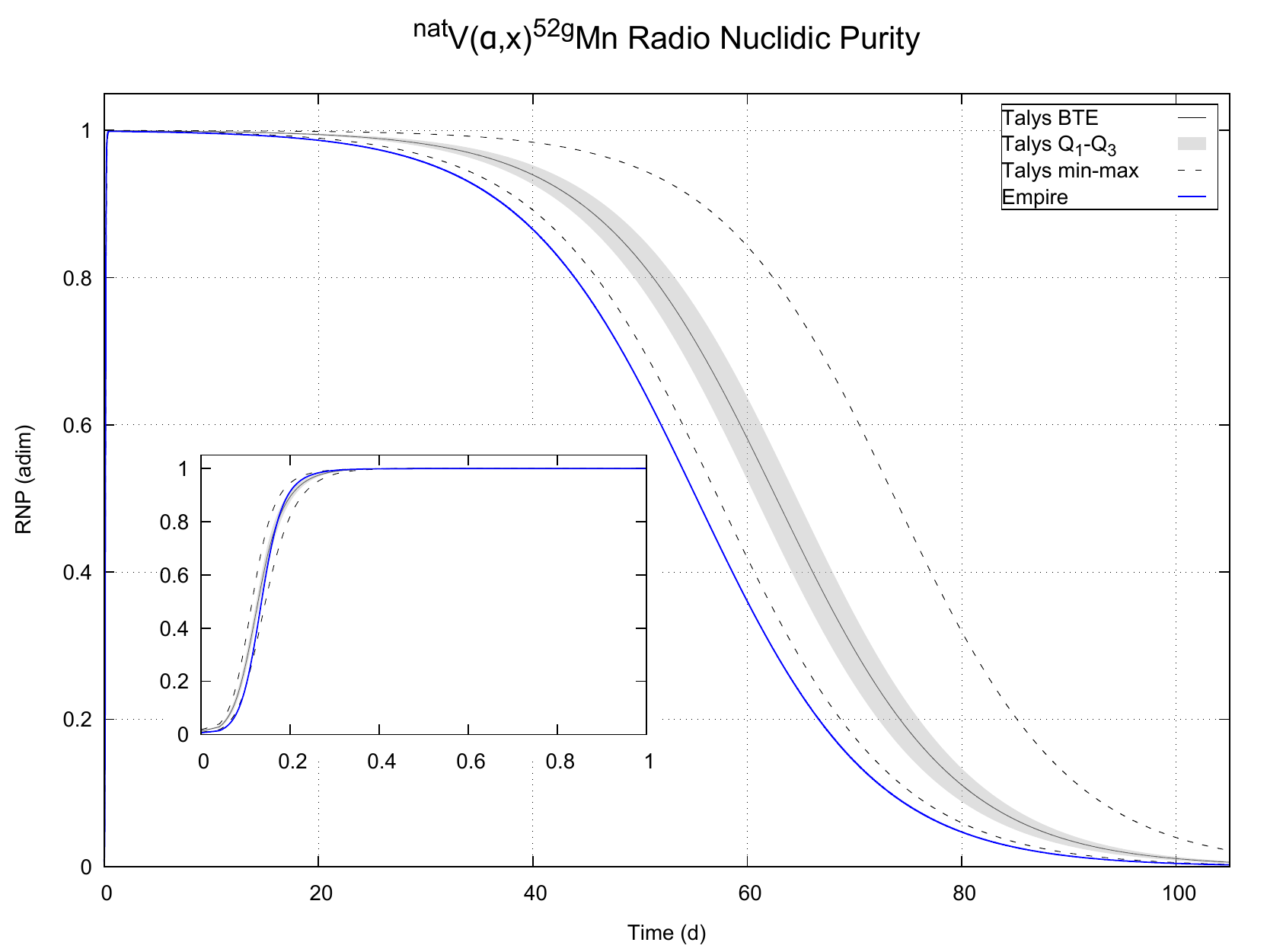}
\caption{Time evolution of the RadioNuclidic Purity (RNP) expected by the considered nuclear models for a long and a short (inset) time window.}
\label{fig:6}
\end{figure}

Finally, the isotopic purity ($IP$) of $^{52g}$Mn may be calculated, as
\begin{equation}
	IP = \frac{n_{^{52g}\mbox{\rm Mn}}}{n_{^{48}\mbox{\rm Mn}}+n_{^{49}\mbox{\rm Mn}}+n_{^{50(g+m)}\mbox{\rm Mn}}+n_{^{51}\mbox{\rm Mn}}+n_{^{52(g+m)}\mbox{\rm Mn}}+n_{^{53}\mbox{\rm Mn}}+n_{^{54}\mbox{\rm Mn}}+n_{^{55}\mbox{\rm Mn}}} \, ,
\end{equation}
where $n$ is the number of nuclei. The corresponding radionuclidic purity ($RNP$) is given by
\begin{equation}
	RNP = \frac{A_{^{52g}\mbox{\rm Mn}}}{A_{^{48}\mbox{\rm Mn}}+A_{^{49}\mbox{\rm Mn}}+A_{^{50(g+m)}\mbox{\rm Mn}}+A_{^{51}\mbox{\rm Mn}}+A_{^{52(g+m)}\mbox{\rm Mn}}+A_{^{53}\mbox{\rm Mn}}+A_{^{54}\mbox{\rm Mn}}} \, ,
\end{equation}
where $A$ represents the activity of the specific isotope. In Figs. \ref{fig:5}-\ref{fig:6} the time evolution of IP and RNP are shown, both for a short and a long time scale.

%However the possibility 
%of high purity $^{52}$Mn %productionfor long times after the %end of the irradiation seem to be 
%predicted by all the models.

At the End of Bombardment (EoB), the IP achieves values of about 0.45-0.75 (Fig. \ref{fig:5}), and the disagreement between the two codes reflects the different results for the cross sections already shown in Fig.\ref{fig:1}.
The value is not so high due to the production of $^{53}$Mn which can be considered stable at the timescale shown, and therefore does not affect significantly the RNP.
Indeed, the RNP achieves a value close to 1 after few hours, due to the very short half-lives of $^{50}$Mn, $^{51}$Mn and $^{52m}$Mn, as well as  to the small production of $^{54}$Mn
and the negligible activity for $^{53}$Mn. 

The Talys BTE value (with the corresponding uncertainty) estimated for the activity of $^{52g}$Mn at EoB is about $6.23 \pm 0.80$ MBq, while the values foreseen by the Empire code gives 3.20 MBq. It is important to observe from Fig. \ref{fig:1} that the experimental cross-section data stand in between the upper limit identified by the Talys BTE calculation and the lower limit characterized by the Empire prediction.
Despite this discrepancy,  the RNP remains however close to one for about 20 days for both calculation tools, as shown in Fig. \ref{fig:6}.
For this reason the nuclear reaction $^{nat}$V($\alpha$,x)$^{52g}$Mn may be considered of particular interest as an alternative route for $^{52g}$Mn production.

\section{Discussion}
\subsection{Comparison with other production routes}

The standard route usually considered for the cyclotron production of $^{52g}$Mn is \newline $^{nat}$Cr(p,x)$^{52g}$Mn and has been recently reviewed in Refs. \cite{Lapy2019,Qaim2020}. 
The first reference reports new data for the cross section and confirms the peak between 12 and 16 MeV, suited  for an hospital-based cyclotron production.
The second reference is a comprehensive and historical review on the production of medical radionuclides and refers to the $^{nat}$Cr(p,x)$^{52g}$Mn reaction as the main production pathway.
Also the deuteron-induced reaction $^{52}$Cr(d,2n)$^{52g}$Mn, with enriched target, has been previously explored as an alternative route and compared with the proton channel \cite{West1987}. 

\begin{figure}[!htb]
\centering
\includegraphics[width=11cm]{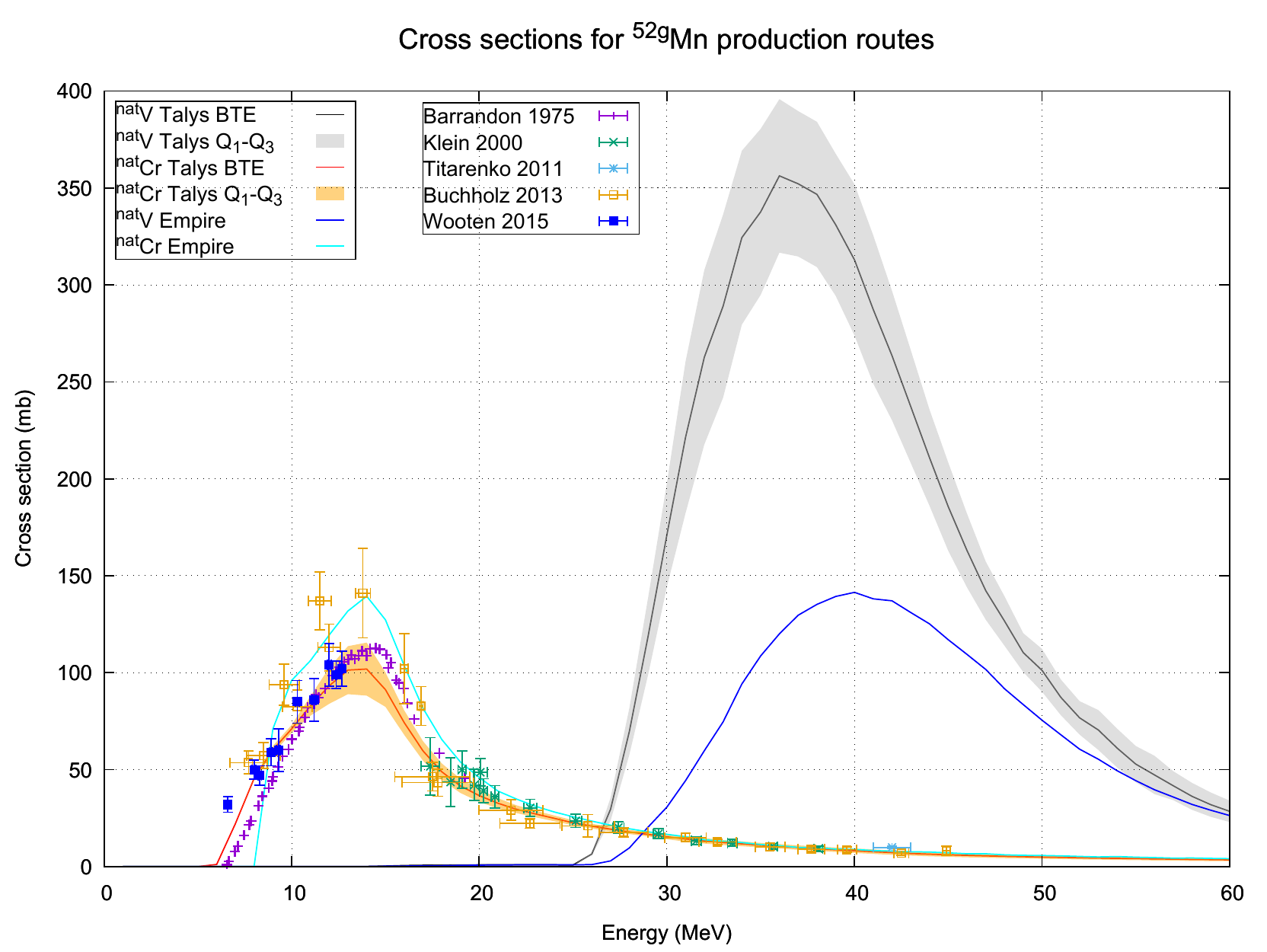}
\caption{Comparison of  $^{52g}$Mn production cross sections from $^{nat}$Cr (proton beams) and $^{nat}$V ($\alpha$ beams) targets. 
Data are shown only for $^{nat}$Cr targets (refer to Fig. \ref{fig:1} for data on $^{nat}$V targets).}
\label{fig:9bis}
\end{figure}

Cross sections, yields, rates and expected radionuclide purities have been evaluated also for reactions involving chromium targets.
In particular, in Fig.\ref{fig:9bis},
the cross sections from $^{nat}$Cr and  $^{nat}$V targets are compared.
In the figure are also plotted the experimental data from EXFOR database for the $^{nat}$Cr(p,x)$^{52g}$Mn reaction
(see Refs. \cite{Barrandon1975}, \cite{Klein2000}, \cite{Titarenko2011}, \cite{Buchholz2013}, \cite{Wooten2015}). Instead, the experimental data for $^{nat}$V($\alpha$,x)$^{52g}$Mn are here not included because they have been already shown in Fig.\ref{fig:1}.

Cross section values for both reactions routes appear comparable in magnitude, and obviously shifted in the energy range.
In case of $^{nat}$Cr targets,
Empire provides good reproduction with a slight overestimation of the peak,  while Talys provides an almost excellent description, with a minimum  underestimation of the peak. The situation with $^{nat}$V targets has been extensively described while commenting Fig.\ref{fig:1}, with the main outcome that Talys probably overestimates significantly the data, the Empire calculations may produce a possible underestimation. However, the experimental data are excessively scattered to get a definitive conclusion.

\begin{figure}[!htb]
\centering
\includegraphics[width=11cm]{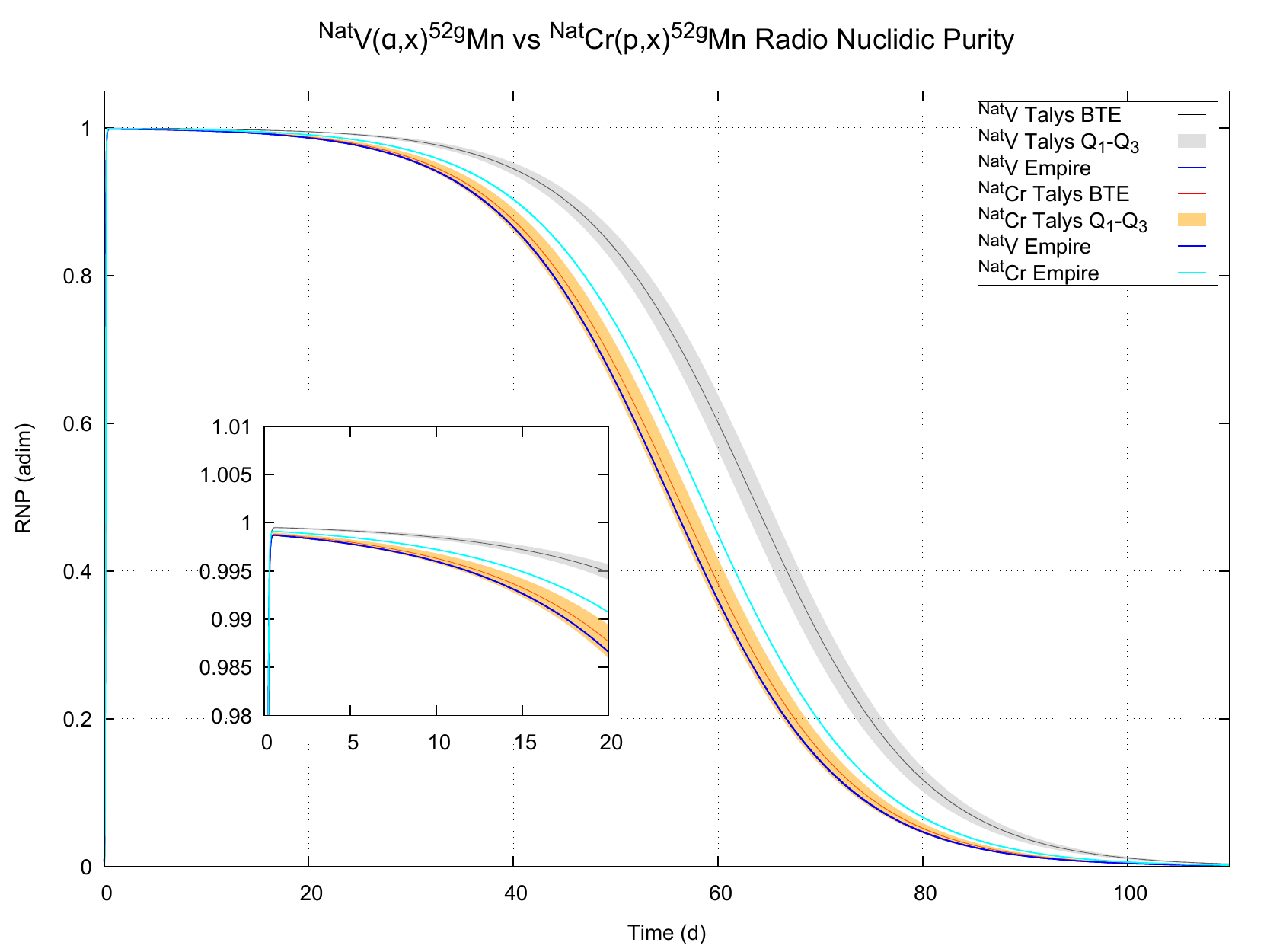}
\caption{Comparison of RNP for $^{nat}$V vs $^{nat}$Cr.}
\label{fig:9}
\end{figure}

It is interesting to compare also the radionuclidic purities that can be obtained with both reactions, as shown in Fig. \ref{fig:9}. The two reaction codes show a different trend, which reflects the different behaviour with the estimated cross section trends. 
Talys highlights a more favorable time evolution of the RNP for $\alpha$ impinging on natural V target. On the contrary, Empire shows a similar behaviour for the two RNPs with a very small advantage for the Chromium targets. However, to draw a definitive conclusion, better cross section experimental data for the reaction $^{nat}$V($\alpha$,x)$^{52g}$Mn are nevertheless needed.

For all the production routes considered, we assess from the integral yields the irradiation energy range corresponding to the highest purities expected. This kind of optimization is crucial because the quantity and quality of the final product are essential for following radiochemistry studies, dosimetric evaluations and, eventually,  pre-clinical studies, to confirm the feasibility of the production route.  

\begin{figure}[!htb]
\centering
\includegraphics[width=12cm]{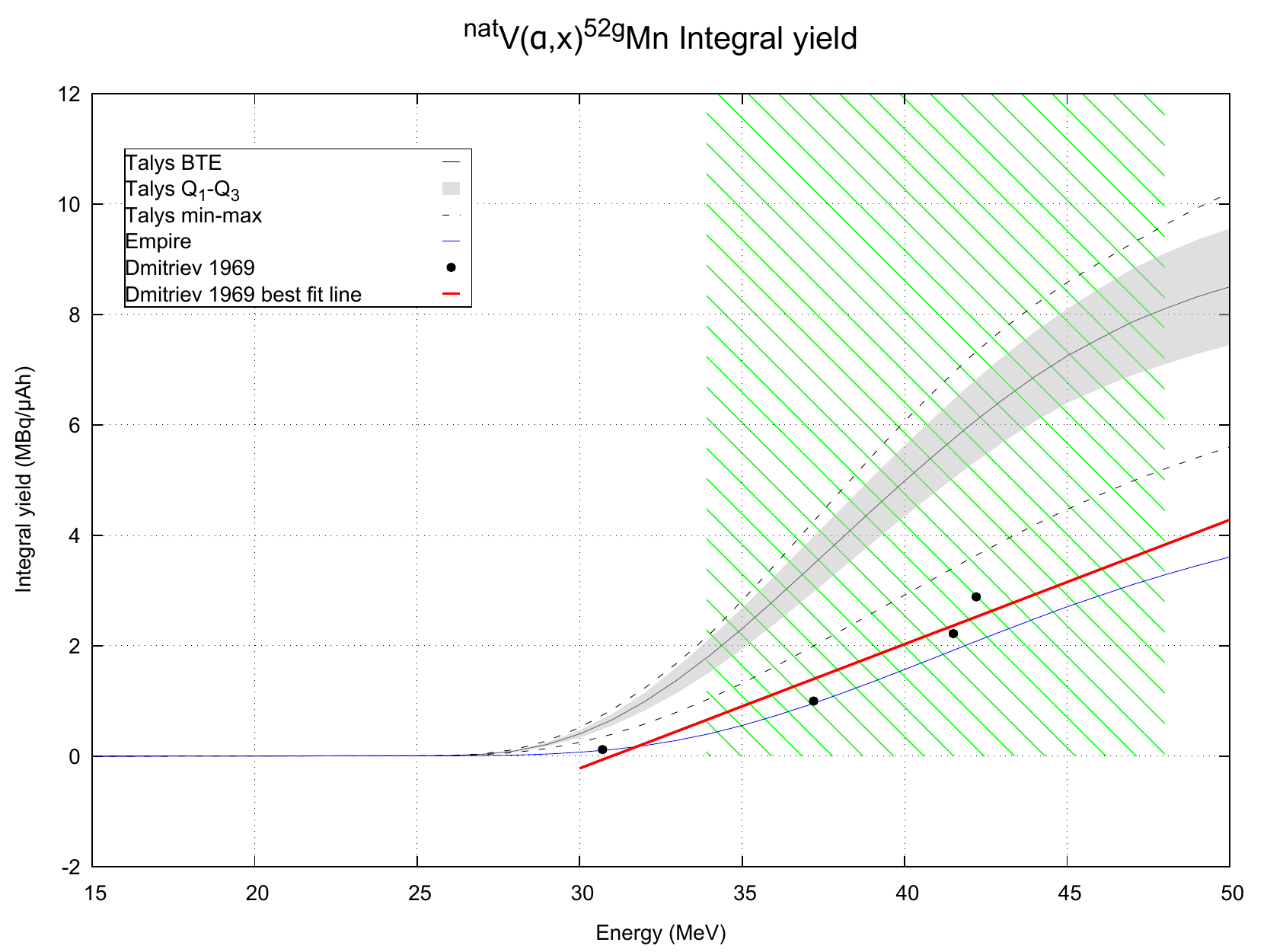}
\caption{$^{52g}$Mn integral yield for a $\alpha$-beam  with 1 $\mu A$ current and one hour irradiation time. The green shaded area indicates the optimized energy interval used for the 200-$\mu$m thick target. }
\label{fig:10}
\end{figure}

\begin{figure}[!htb]
\centering
\includegraphics[width=12cm]{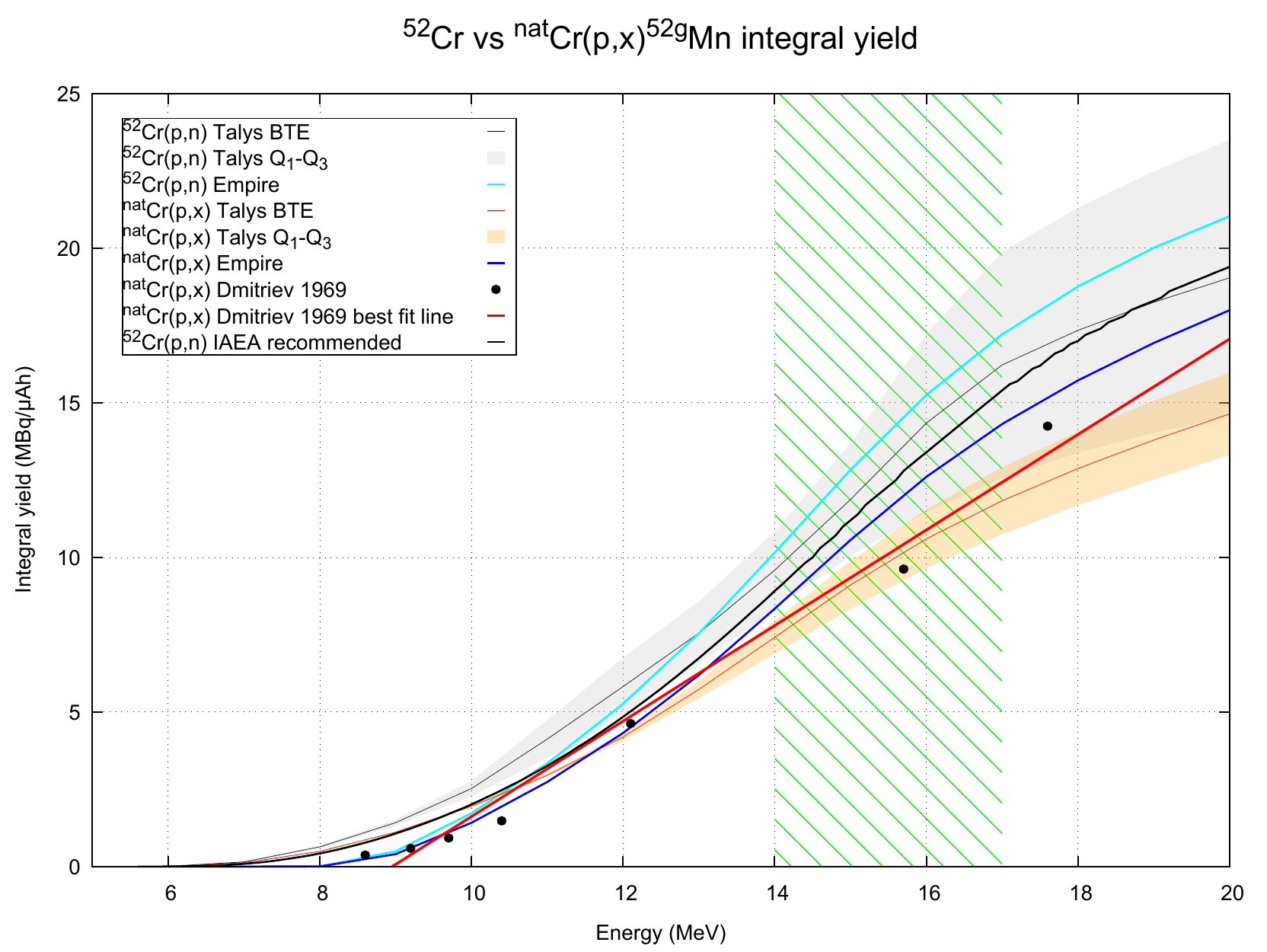}
\caption{$^{52g}$Mn integral yield from $^{nat}$Cr and enriched $^{52}$Cr targets for a proton-beam with 1 $\mu A$ current and one hour irradiation time. The green shaded area indicates the optimized energy interval used for the 200-$\mu$m thick target.}
\label{fig:12}
\end{figure}

\begin{figure}[!htb]
\centering
\includegraphics[width=12cm]{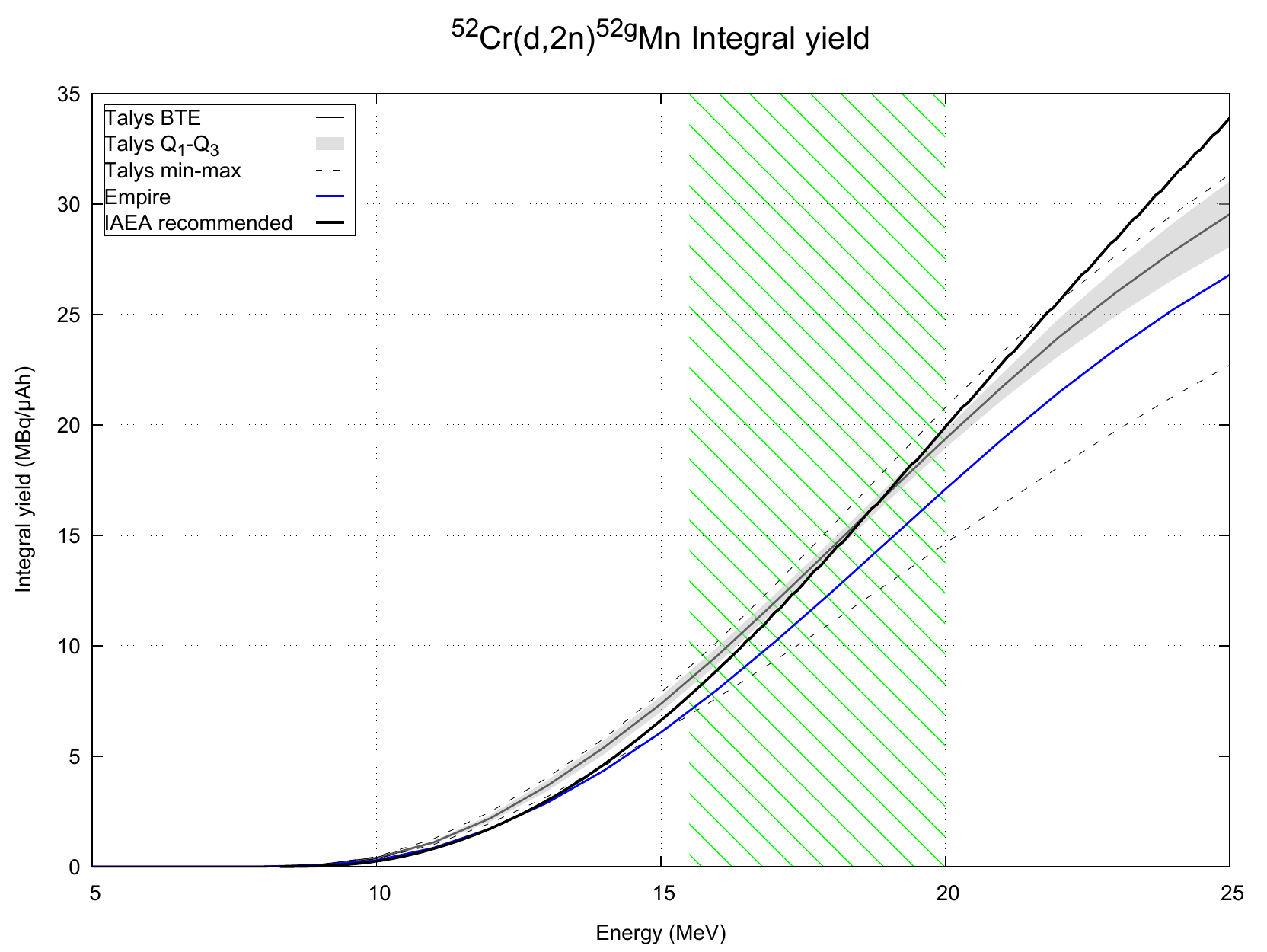}
\caption{$^{52g}$Mn integral yield for a deuteron-beam irradiation on enriched $^{52}$Cr. The irradiaton conditions are the same of Figs. \ref{fig:10} and \ref{fig:12}.
 }
\label{fig:11}
\end{figure}

In Figs. \ref{fig:10} and \ref{fig:12} the integral yields obtained for the reactions $^{nat}$V($\alpha$,x)$^{52g}$Mn and\newline $^{nat}$Cr(p,x)$^{52g}$Mn~are~respectively shown. For comparison, in the latter figure the integral yield obtained with the enriched chromium target,
$^{52}$Cr(p,n)$^{52g}$Mn has also been added and finally, the analysis has been completed by considering also the production channel $^{52}$Cr(d,2n)$^{52g}$Mn, shown in Fig. \ref{fig:11}.

In all cases, the reference irradiation parameters assumed in this work to estimate the integral yields are:
1h irradiation time; 1 $\mu A$ for the beam current; and a target thickness large enough to stop completely the beam inside the target. Curves show the variation of the integral yield against the incident beam energy, and represents the production yield for a hypothetical beam stopper target. From these curves it is possible to draw the yield for a target of a given thickness by taking the difference between the values corresponding to the ingoing and outgoing energies (values delimited by the green shaded area).
For all targets a thickness of 200 $\mu m$ has been assumed and the beam output energy of such target, E$_o$, has been derived from the Bethe-Bloch equation \cite{leo}. The yield of a target with 200 $\mu m$ thickness is then given by the difference of the integral yield evaluated at E$_i$ and  E$_o$, where E$_i$ denotes the incoming-beam energy. 

In all cases concerned, the beam energy window has been optimized taking into account the steepness of the integral yield, the minimization in the contaminant production and the 200 $\mu$m constraint of the target thickness. The resulting energy ranges are reported in the left column of Tab. \ref{tab:3}, and highlighted as well by a vertical green shaded area in Figs. \ref{fig:10},  \ref{fig:12}, and \ref{fig:11}.

The results with $^{nat}$V are given
in Fig. \ref{fig:10} with the integral yields obtained by Talys and Empire codes. For Talys the results are given in terms of the BTE value and its associated uncertainty, following Eq. \ref{eq:1}, which is highlighted as a gray band on figure. In addition, the data by Dmitriev at al. \cite{Dmitriev1969} are reported as well, along  with a linear interpolation obtained from these data. 
As it is clearly shown, it appears that Talys overestimates the yield by a factor of about 2, in line 
with our findings with the cross sections (see Fig.\ref{fig:1}). On the other hand, the experimental data (and/or its linear interpolation) and Empire calculations appear fairly consistent.


In Fig. \ref{fig:12}  the integral yields for a proton beam on 
$^{nat}$Cr as well as on  100\% enriched $^{52}$Cr targets are compared. The results are similar with the enriched target case overperforming with respect to the natural one. The expected results from Talys are very close to the IAEA recommended yields \cite{livechart} for the enriched chromium target, and similarly, for the natural target, they are close to the data interpolation. In both cases, Empire gives similar results, with a slight yield overestimation.

In Fig. \ref{fig:11} the case of deuteron beam on enriched $^{52}$Cr target is plotted. In the range of interest, [20-15.5] MeV, Talys and the IAEA recommended values are very close, with the IAEA recommended curve slightly steeper, while Empire results are still close although somewhat lower.

The analysis performed in Figs. \ref{fig:10}, \ref{fig:12}, and \ref{fig:11} allows to draw the production yields of a target of given thickness, for instance the benchmark 200 $\mu$m assumed in our comparative study. Such quantity is readily calculated as the difference of the plotted integral yields at the bombarding energy E$_i$, and at the target exit energy E$_o$, which takes into account the energy loss in the material.  Tab. \ref{tab:3} compares the derived $^{52g}$Mn production yield determined by the nuclear reaction codes, and when available, from experimental measures \cite{Dmitriev1969}, as well as from nuclear data evaluations \cite{livechart}.
We have added in the comparison also the output obtained by the ARRONAX Radionuclide Yield Calculator (RYC) \cite{ryc}, based on the TENDL library \cite{TENDL}.

It is difficult to recommend a value for the first route reported in Tab. \ref{tab:3} and this reflects the large experimental uncertainty  obtained for the cross section, as reported in Fig. \ref{fig:1}. Clearly Talys evaluations and the similar RYC value overestimate the yield, while Empire probably underestimates it.
One can tentatively extract a significant guess by observing that the Empire curve reproduces the overall trend of the cross section data, if it is rescaled by a factor 1.5/1.7. Under this assumption, we can accordingly rescale the Empire yield in Tab. \ref{tab:3} obtaining approximately 4.3/4.9 MBq/$\mu$Ah. This crude "guesstimate" compares well with the production yield of 4.4 calculated with Talys for protons on $^{nat}$Cr target (see the second line in the Table). As one can see from Fig.\ref{fig:9bis}, for this reaction the Talys results are quite reliable and the corresponding yield provides a good estimate. Therefore, we may conjecture that the production yield for the two routes could be very similar, although a definitive conclusion can be drawn only with much better data for the $^{nat}$V($\alpha$,x)$^{52g}$Mn cross section. Alternatively, a careful nuclear data evaluation of the existing measurements for this cross section is strongly needed. The comparison in Tab. \ref{tab:3} is completed with two $^{52}$Cr-target reactions, the first with proton and the second with deuteron beams. These two routes are well known and listed in the IAEA medical radioisotopes production database \cite{livechart} with recommended evaluations for the cross sections. The use of enriched material in the $^{52}$Cr(p,n) route provides an additional 15\% production yield in comparison to $^{nat}$Cr, with the advantage of a drastic reduction of contaminants, as will be shown in Tab. \ref{tab:3bis}. The $^{52}$Cr(d,2n) route produces even more $^{52g}$Mn (almost a 100\% addition), but here the contaminant reduction is not so efficient. 

The production of the most relevant contaminants derived under the same irradiation conditions are given in Tab.\ref{tab:3bis}. By far the most critical radionuclide is $^{54}$Mn.
In the case of $\alpha$ particles colliding on $^{nat}$V the cross section peaks around 12/14 MeV and rapidly decreases toward very low values at higher energies, in particular in the energy interval of our interest. By comparing data, the yield of this contaminant is significantly lower by using $^{nat}$V targets than for $^{nat}$Cr ones. This is a clear advantage when comparing natural targets. Obviously, the use of about 100\%
enriched targets allows to drastically reduce or completely remove the level of contamination, as shown in the third and fourth line of Tab. \ref{tab:3bis}. 

The effect of the contamination  by $^{53}$Mn on the total dose released to the patient has not been carefully analyzed yet \cite{brandt2019}, but it should be of minor importance due to its very long half-life. The very small activities reported in Tab. \ref{tab:3bis} for this radionuclide
strengthen this assumption. 
Nevertheless the $^{nat}$V route produces a slightly larger yield than the $^{nat}$Cr one, but still small and very close to each other. 
For the enriched target cases the route via the reaction $^{52}$Cr(d,n)$^{53}$Mn gives yield comparable to those with natural targets, while the  $^{52}$Cr(p,$\gamma$)$^{53}$Mn
reaction gives even smaller yield.

\begin{table}[!htb]
\renewcommand*{\arraystretch}{1.5}
\caption{{Comparison of the four production routes analyzed. The irradiation parameters correspond to 1 $\mu$A current and 1 h irradiation time. The optimized energy windows for each route, shown in the left column, correspond to a 200 $\mu$m target thickness. We report Talys calculations with a theoretical error evaluation depending on the variability of the models.} }
\label{tab:3}
\centering
\begin{tabular}{|c|c|c|c|c|}
\hline
Reaction [E$_{i}$-E$_{o}]$ (MeV)& \multicolumn{4}{c|} {Yield (MBq/$\mu$Ah)} \\
\cline{2-5}
 & Talys  & Empire & RYC & Data fit or \\
 & &  &  & interpolation  \\
\hline
$^{nat}$V($\alpha$,x)$^{52g}$Mn  
[48-33.9]
& 6.28 $\pm$ 1.27 & 2.88 & 5.57  & 3.17 \cite{Dmitriev1969} \\
$^{nat}$Cr(p,x)$^{52g}$Mn [17-14]& 4.41 $\pm$ 0.51  & 5.98 & 4.28 & 5.52 \cite{Dmitriev1969} \\ 
$^{52}$Cr(p,n)$^{52g}$Mn\,\,\, [17-14] & 6.64 $\pm$ 1.73 & 7.06 & 4.75 & 6.47 \cite{livechart} \\
$^{52}$Cr(d,2n)$^{52g}$Mn [20-15.5] & 12.00 $\pm$ 0.63 & 10.09 & 14.43 & 12.14 \cite{livechart} \\
\hline 
\end{tabular}
\end{table}

\begin{table}[!htb]
\renewcommand*{\arraystretch}{1.5}
\caption{{Yields for the main contaminants,  $^{54}$Mn and $^{53}$Mn, with the same irradiation conditions discussed in Tab. \ref{tab:3}.}}
\label{tab:3bis}
\centering
\begin{tabular}{|c|c|c|c|c|}
\hline
Contaminants& \multicolumn{4}{c|} {Yield} \\ 
\cline{2-5}
 & Talys  & Empire & RYC & Unit \\ \hline
$^{nat}$V($\alpha$,x)$^{54}$Mn & 1.94 $\pm$ 0.16 & 3.35 & 2.65 & KBq/$\mu$Ah \\
$^{nat}$Cr(p,x)$^{54}$Mn & 4.80 $\pm$ 0.07  & 5.08 & 3.80 & \\
$^{52}$Cr(p,x)$^{54}$Mn & - & - & - &  \\ 
$^{52}$Cr(d,$\gamma$)$^{54}$Mn & 7.52 $\pm$ 0.92 & 41.8 & 6.7 & Bq/$\mu$Ah \\ \hline 
$^{nat}$V($\alpha$,x)$^{53}$Mn & 10.6 $\pm$ 1.03 & 12.9 & 7.27 & mBq/$\mu$Ah \\
$^{nat}$Cr(p,x)$^{53}$Mn & 7.52 $\pm$ 0.24  & 7.13 & 6.83 & \\
$^{52}$Cr(p,$\gamma$)$^{53}$Mn & 0.23 $\pm$ 0.04 & 0.03 & 0.26 & \\ 
$^{52}$Cr(d,n)$^{53}$Mn & 9.90 $\pm$ 0.24 & 1.44 &7.15 &  \\ \hline
\end{tabular}

\end{table}

\section{Conclusion}

The $^{nat}$V($\alpha$,x)$^{52g}$Mn nuclear reaction route, as a viable alternative for the production of 
the radionuclide $^{52g}$Mn, has been investigated in the present work. This radionuclide is of significant medical interest for the innovative PET-MRI multi-modal imaging technique. This uncommon reaction route has not been considered so far for the production of the radionuclide concerned and does not often appear in the relevant literature.
We have compared this production method with the well known approach via low-energy protons on $^{nat}$Cr targets. The study has been completed by also considering enriched $^{52}$Cr targets bombarded both with proton and deuteron beams.

The experimental data so far available on world databases for $^{nat}$V($\alpha$,x)$^{52g}$Mn appear with a remarkable spread, thus preventing a precise determination for the cross section values. All Talys models that we have included in the calculations provide an overestimation of the peak behaviour around 40 MeV, while the Empire results somewhat underestimates the cross section. Although the nuclear model tools used in the present work do not describe the specific reaction for $^{52g}$Mn in an optimal way, they provide at least an upper (Talys) and lower (Empire) bound that delimit the measurements. 
The cross sections for all other Mn contaminants are in better agreement and have been considered to find out a favorable production energy window, in terms of yields, isotopic and radionuclidic purities.

Our study shows that, even if we consider the most conservative estimation from Empire, the production yield is significant to make $^{nat}$V($\alpha$,x)$^{52g}$Mn of interest. In addition, concerning the production of the main contaminant, $^{54}$Mn, the reaction provides a better product in terms of purity with respect to $^{nat}$Cr(p,x)$^{52g}$Mn. One must acknowledge the fact that with $^{nat}$Cr the production can be achieved with hospital cyclotrons exploiting low-energy proton beams, while the $^{nat}$V production requires a 50 MeV cyclotron and $\alpha$ particles, which can be currently found only in few research centers. Nevertheless, for infrastructures where this kind of machines are available, it might be convenient to consider this alternative reaction.

The $^{52g}$Mn production based upon enriched $^{52}$Cr targets presents significant advantages in production yields and quality, due to minor presence of Mn contaminants, but on the other hand it requires the use of more expensive materials and specific technologies for target recovery. 

\section*{Declarations}

\subsection*{Ethics approval and consent to participate}
  Not applicable.
\subsection*{Consent for publication}
  Not applicable.
\subsection*{Availability of data and material}
  All experimental data used for this work have been referenced. Contact the corresponding authors for material and data presented in this work.
\subsection*{Competing interests}
  The authors declare that they have no competing interests.
\subsection*{Funding}
  This research was carried out within the METRICS applied physics program (2018–2020/2021), approved and founded by the INFN--CSN5 (Technological Research Committee).
\subsection*{Author's contributions}
   All authors contributed equally to the work presented.
\subsection*{Acknowledgements}
  The authors are grateful to Arjan Koning, Roberto Capote, Paola Sala, and Alfredo Ferrari,  for stimulating discussions and insights about the use of the nuclear reaction codes. Discussions with Laura De Nardo and Juan Esposito on the research subject are thankfully acknowledged. 

 \section*{Appendix}
\begin{figure}[!htb]
\centering
\includegraphics[width=12cm]{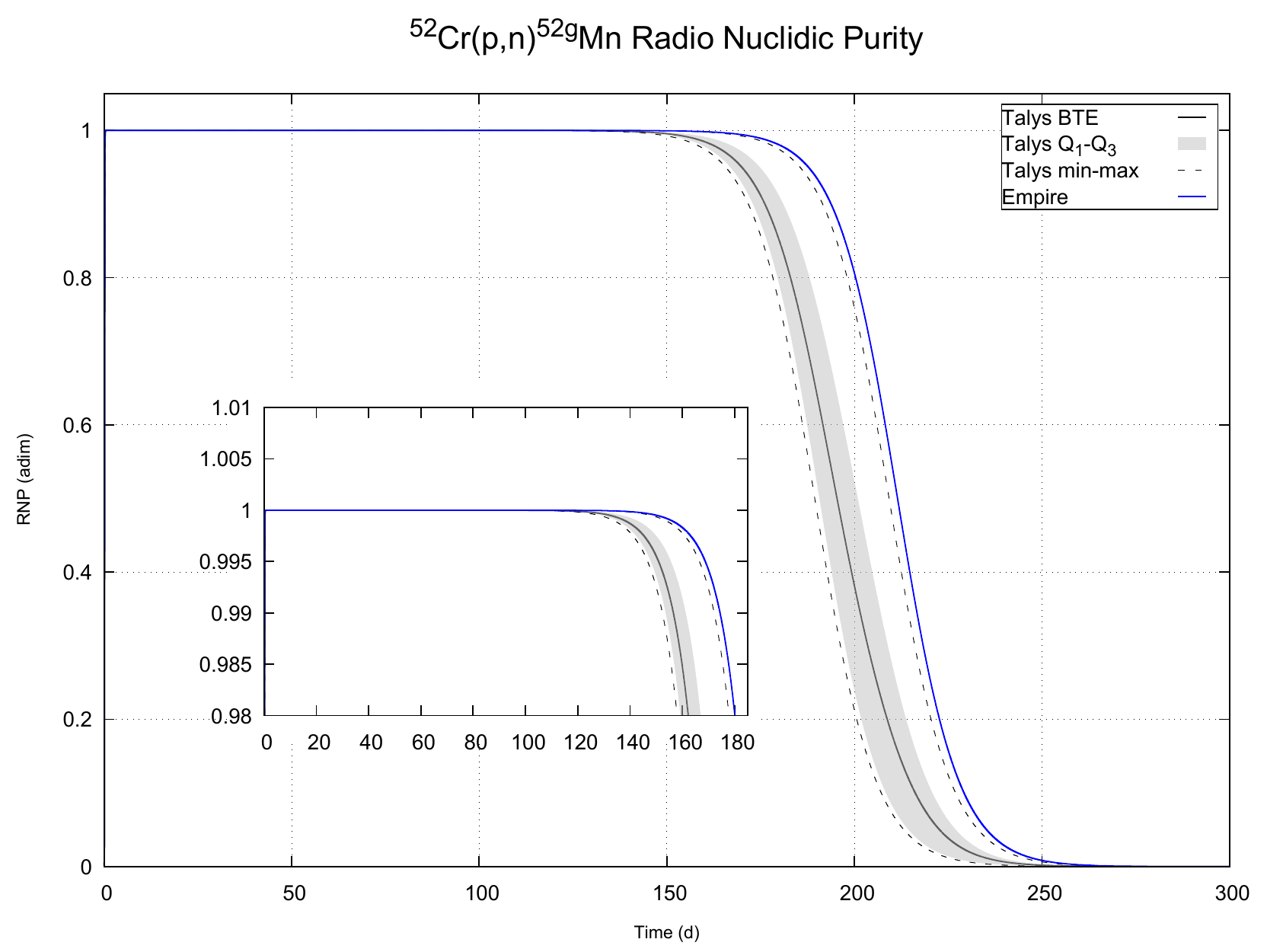}
\caption{Time evolution of the radionuclidic purity for the reaction
$^{52}$Cr(p,n)$^{52g}$Mn assuming an 100\% target enrichment. The irradiation conditions are those discussed for Tab.\ref{tab:3}.
}
\label{fig:7}
\end{figure}

\begin{figure}[!htb]
\centering
\includegraphics[width=12cm]{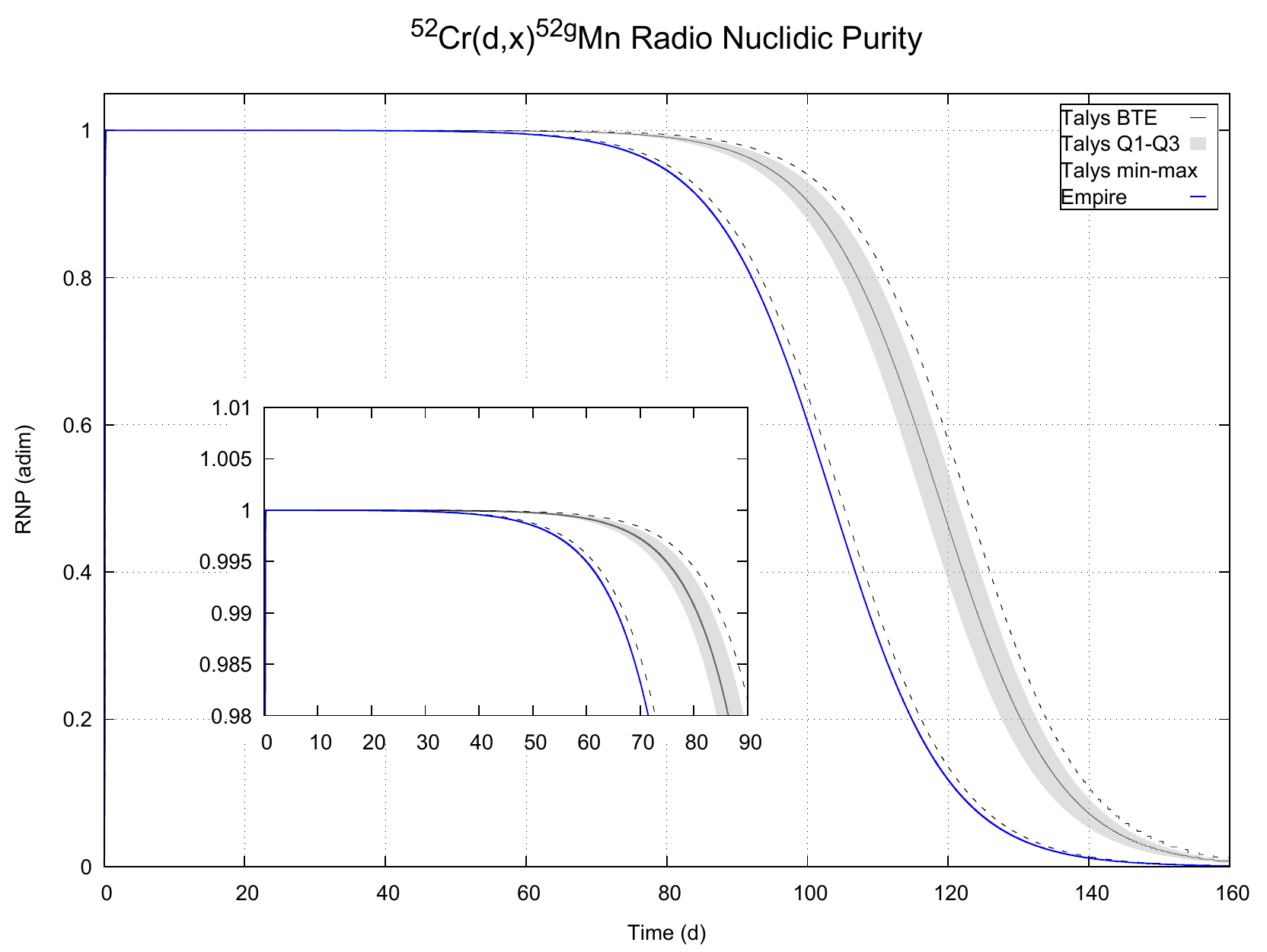}
\caption{The same as Fig. \ref{fig:7} for the reaction $^{52}$Cr(d,2n)$^{52g}$Mn.}
\label{fig:8}
\end{figure}

 We focus here on how the enrichment of both chromium and vanadium targets affects the production of $^{52g}$Mn and contaminants.
 Specifically, we illustrate in Figs. \ref{fig:7} and \ref{fig:8}, respectively, the time evolution of RNP with proton and deuteron beams impinging on $^{52}$Cr target with hypothetical 
100\% enrichment.
 
 If we compare these two figures with the corresponding RNP obtained with natural targets, Fig. \ref{fig:9}, it is evident that very high purity levels can be maintained over a much longer time in case of enriched targets. With natural targets, more than 20 days are needed before RNP reduces by 1.5 \%. Much higher values are needed for protons and deuterons on enriched 
targets, 150 and 65 days, respectively. 
However, considering the 5.6 d half-life of $^{52g}$Mn, the result with natural targets appears adequate to maintain a sufficiently high RNP, at least for more than three half-lives.

\begin{figure}[!htb]
\centering
\includegraphics[width=12cm]{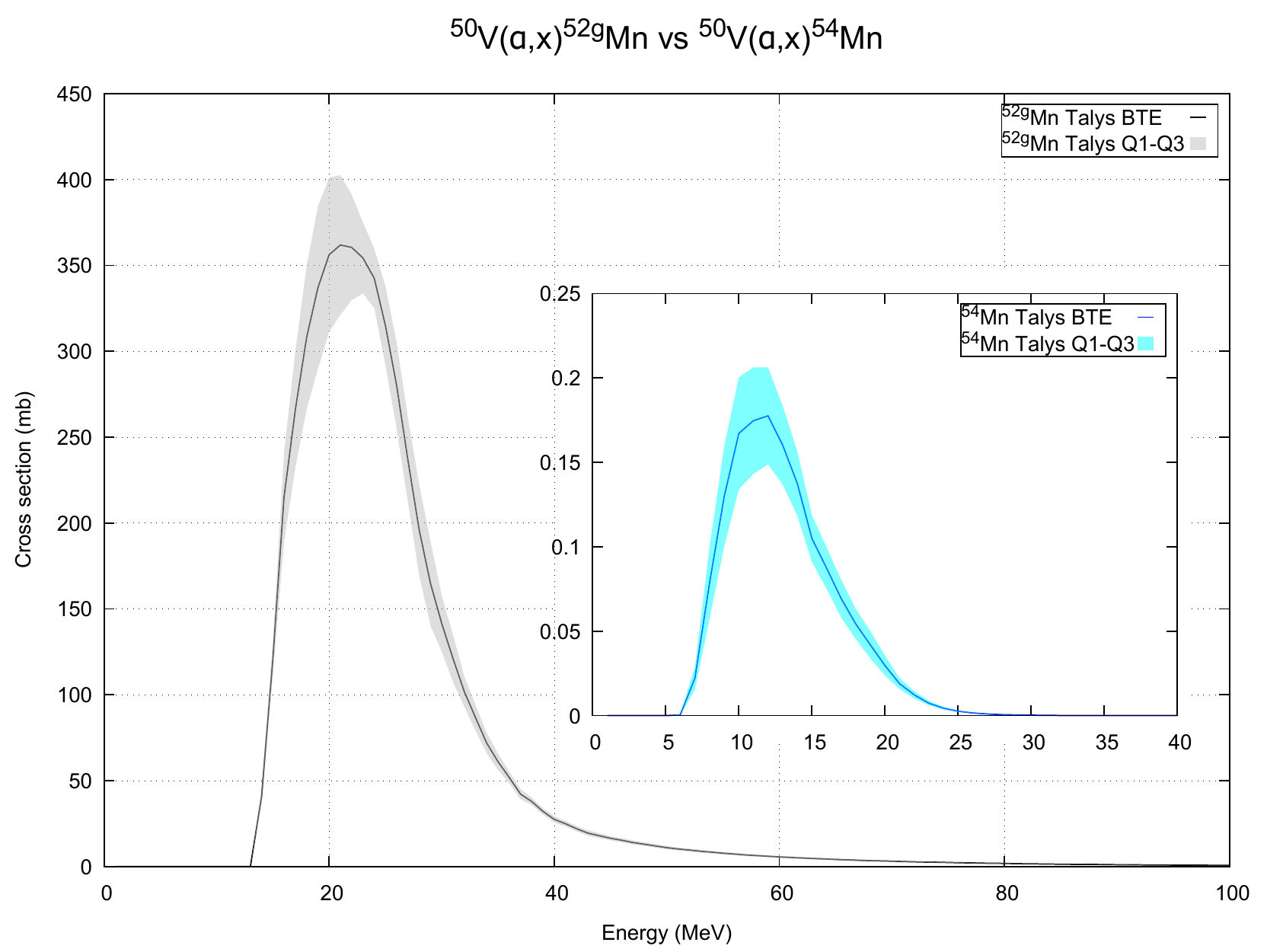}
\caption{Cross sections for the production of $^{52g}$Mn and $^{54}$Mn with alpha beams on $^{50}$V.}
\label{fig:Blin}
\end{figure}

\begin{figure}[!htb]
\centering
\includegraphics[width=12cm]{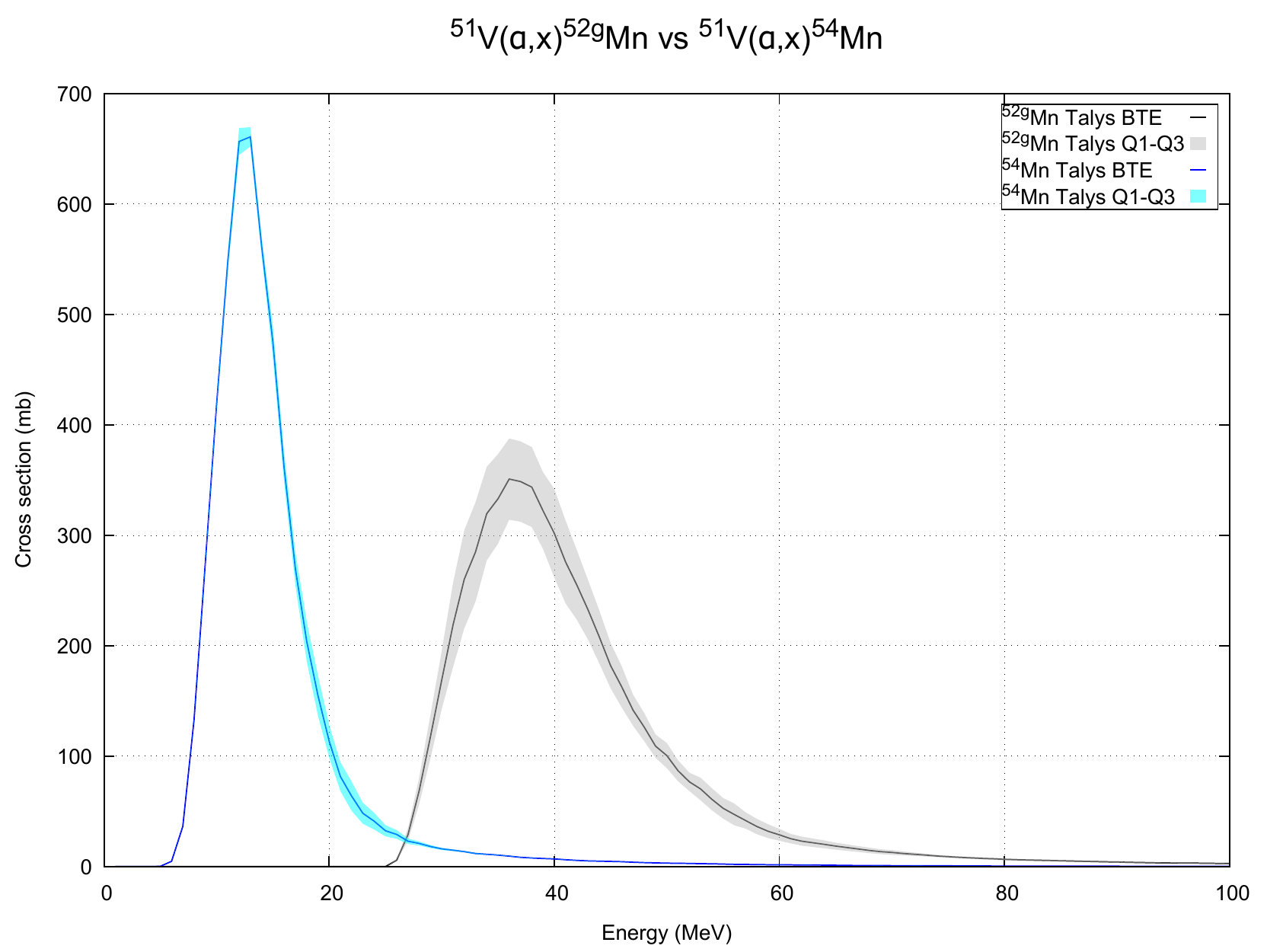}
\caption{Cross sections for the production of $^{52g}$Mn and $^{54}$Mn with alpha beams on $^{51}$V.}
\label{fig:A}
\end{figure}

Vanadium has to be considered a monoisotopic element made of $^{51}$V, but the presence in small fraction (0.25\%)  of the radioactive $^{50}$V, with 1.5 $\times$ 10$^{17}$ y  half-life, does not make it mononuclidic. It might be odd, but since it is possible to find commercially samples of vanadium  with $^{51}$V abundance different from $^{nat}$V, we discuss how this could affect the production. In Figs. \ref{fig:Blin} and  \ref{fig:A}  the cross sections for $^{50}$V($\alpha$,x)$^{52g,54}$Mn  and $^{51}$V($\alpha$,x)$^{52g,54}$Mn 
provide a clear representation of the reaction dynamics at stake, and exhibit 
the fine balance between production of the radionuclide of interest and its main contaminant.
Following the scheme previously adopted, the calculations have been performed by evaluation of the Talys interquartile range and corresponding BTE.

Radioactive $^{50}$V has the advantage of a minimum production of $^{54}$Mn contaminant,  and implies a  $^{52g}$Mn peak around 20 MeV, at significantly lower energies than the peak with $^{51}$V. These features could be attractive in an ideal and very hypothetical  situation of about 100\% $^{50}$V target, but become an issue in the case of a $^{51}$V compresence. Indeed,  the $^{52g}$Mn peak shifts at lower energies 
with increasing abundance of $^{50}$V, and this interferes with the $^{54}$Mn production from $^{51}$V, which significantly increases at lower energies as well. On the other hand, there is no appreciable advantage when considering enriched $^{51}$V targets, either. A $^{51}$V target with 100\% enrichment does not improve the $^{52g}$Mn production nor reduce the contaminant production.

\pagebreak
\bibliographystyle{style/ans_js}                                                                           %custom ANS journal submission template bibliography style
\bibliography{vnata}

\end{document}